**A novel technique for image steganography based on Block-DCT and Huffman Encoding**

A.Nag[!], S. Biswas[*], D. Sarkar[*], P.P. Sarkar[*]

[!]Academy of Technology, Hoogly – 721212
e-mail: it_amitava@yahoo.co.in

*USIC, University of Kalyani, Kalyani, Nadia – 741 235, West Bengal, India.
e-mail: mirror@sancharnet.in

***Abstract:*** *Image steganography is the art of hiding information into a cover image. This paper presents a novel technique for Image steganography based on Block-DCT, where DCT is used to transform original image (cover image) blocks from spatial domain to frequency domain. Firstly a gray level image of size M × N is divided into no joint 8 × 8 blocks and a two dimensional Discrete Cosine Transform(2-d DCT) is performed on each of the P = MN / 64 blocks. Then Huffman encoding is also performed on the secret messages/images before embedding and each bit of Huffman code of secret message/image is embedded in the frequency domain by altering the least significant bit of each of the DCT coefficients of cover image blocks. The experimental results show that the algorithm has a high capacity and a good invisibility. Moreover PSNR of cover image with stego-image shows the better results in comparison with other existing steganography approaches. Furthermore, satisfactory security is maintained since the secret message/image cannot be extracted without knowing decoding rules and Huffman table.*

## 1. INTRODUCTION

With the development of Internet technologies, digital media can be transmitted conveniently over the Internet. However, message transmissions over the Internet still have to face all kinds of security problems. Therefore, how to protect secret messages during transmission becomes an essential issue for the Internet. Encryption is a well-known procedure for secure data transmission. The commonly used encryption schemes include DES (Data Encryption Standard) [1], AES (Advanced Encryption Standard)[2] and RSA [3]. These methods scramble the secret message so that it cannot be understood. However, it makes the message suspicious enough to attract eavesdropper's attention. Hence, a new scheme, called "steganography" [4], arises to conceal the secret messages within some other ordinary media (i.e. images, music and video files) so that it cannot be observed. Steganography differs from cryptography in the sense that where Cryptography focuses on concealing the contents of a message, steganography focuses on concealing the existence of a message [5].

Two other technologies that are closely related to steganography are watermarking and fingerprinting [6].Watermarking is a protecting technique which protects (claims) the owner's property right for digital media (i.e. images, music, video and software) by some hidden watermarks. Therefore, the goal of steganography is the secret messages while the goal of watermarking is the cover object itself.

Steganography is the art and science of hiding information in a cover document such as digital images in a way that conceals the existence of hidden data. The word steganography in Greek means "covered writing" ( Greek words "*stegos*" meaning "cover" and "*grafia*" meaning "writing") [7]. The main objective of steganography is to communicate securely in such a way that the true message is not visible to the observer. That is unwanted parties should not be able to distinguish in any sense between cover-image (image not containing any secret message) and stego-image (modified cover-image that containing secret message). Thus the stego-image should not deviate much from original cover-image. Today steganography is mostly used on computers with digital data being the carriers and networks being the high speed delivery channels. Figure. 1 shows the block diagram of a simple steganographic system.

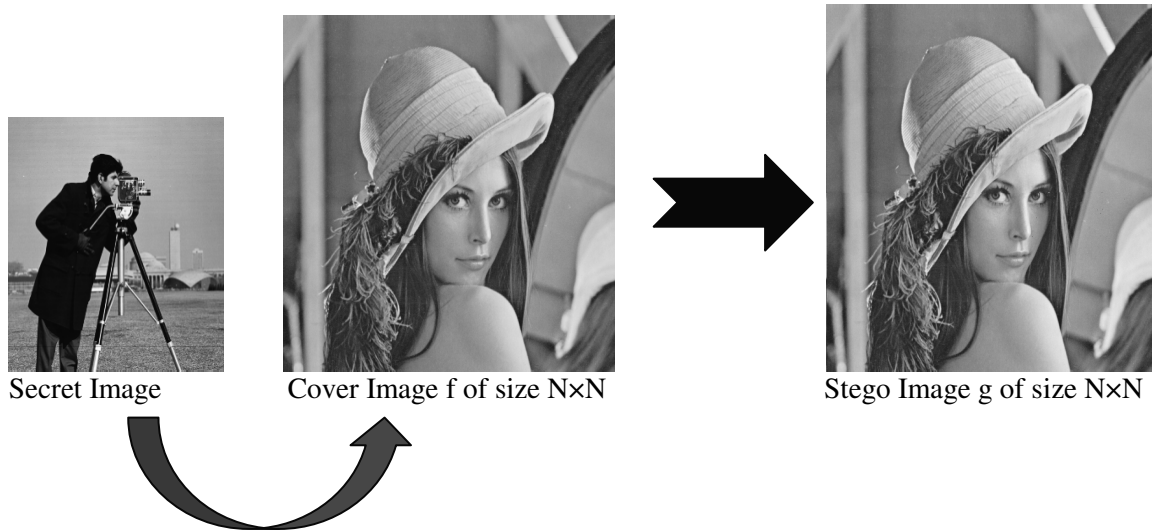

**Figure. 1** The block diagram of a simple steganographic system

## 1.1. RELATED WORK

Information hiding is an old but interesting technology [8]. Steganography is a branch of information hiding in which secret information is camouflaged within other information. A simple way of steganography is based on modifying the least significant bit layer of images, known as the *LSB technique* [9]. The LSB technique directly embed the secret data within the pixels of the cover image. In some cases (Fridrich et al. [10]) LSB of pixels visited in random or in certain areas of image and sometimes increment or decrement the pixel value. Some of the recent research studied the nature of the stego and suggested new methodologies for increasing the capacity. Habes in [11] proposed a new method (4 least Significant) for hiding secret image inside carrier image. In this method each of individual pixels in an image is made up of a string of bits. He took the 4-least significant bit of 8-bit true color image to hold 4-bit of the secret message /image by simply overwriting the data that was already there.

The schemes of the second kind embed the secret data within the cover image that has been transformed such as DCT (discrete cosine transformation). The DCT transforms a cover image from an image representation into a frequency representation, by grouping the pixels into non-overlapping blocks of $8 \times 8$ pixels and transforming the pixel blocks into 64 DCT coefficients each [12,13,14]. A modification of a single DCT coefficient will affect all 64 image pixels in that block. The DCT coefficients of the transformed cover image will be quantized, and then modified according to the secret data. Tseng and Chang in [14] proposed a novel steganography method based on JPEG. The DCT for each block of 8×8 pixels was applied in order to improve the capacity and control the compression ratio.

Capacity, security and robustness [16], are the three main aspects affecting steganography and its usefulness. Capacity refers to the amount of data bits that can be hidden in the cover medium. Security relates to the ability of an eavesdropper to figure the hidden information easily. Robustness is concerned about the resist possibility of modifying or destroying the unseen data.

### 1.2 PSNR (Peak Signal to Noise Ratio)

The measurement of the quality between the cover image f and stego-image g of sizes N × N shown in figure 1 is defined as:

$$PSNR = 10 \times \log(255^2 / MSE)$$

$$\text{where } MSE = \sum_{x=0}^{N-1} \sum_{y=0}^{N-1} (f(x, y) - g(x, y))^2 / N^2$$

Where f(x,y) and g(x,y) means the pixel value at the at position (x, y) in the cover-image and the corresponding stego-image respectively. The PSNR is expressed in dB's. The larger PSNR indicates the higher the image quality i.e. there is only little difference between the cover-image and the stego-image. On the other hand, a smaller PSNR means there is huge distortion between the cover-image and the stego-image.

## 2. PROPOSED IMAGE STEGANOGRAPHY ALGORITHM

Image steganography schemes can be classified into two broad categories: spatial-domain [17, 18,19] based and transform-domain based [20, 21, 22]. In spatial domain approaches, the secret messages are embedded directly. On spatial domain, the most common and simplest steganographic method [24, 9] is the least significant bits (LSB) insertion method. In the LSB technique, the least significant bits of the pixels is replaced by the message which bits are permuted before embedding. However, the LSB insertion method is easy to be attacked. In [23], a new steganography technique, named, "modified side match scheme" was proposed. It reserves the image quality, increases embedding capacity but is not robust against attack because it is a spatial domain approach and no transfer is used. Based on the same embedding capacity, our proposed method improves both image quality and security.

Hiding the secret message/image in the special domain can easily be extracted by unauthorized user. In this paper, we proposed a frequency domain steganography technique for hiding a large amount of data with high security, a good invisibility and no loss of secret message. The basic idea to hide information in the frequency domain is to alter the magnitude of all of the DCT coefficients of cover image. The 2-D DCT convert the image blocks from spatial domain to frequency domain. The schematic/ block diagram of the whole process is given in figure 2((a) and (b)).

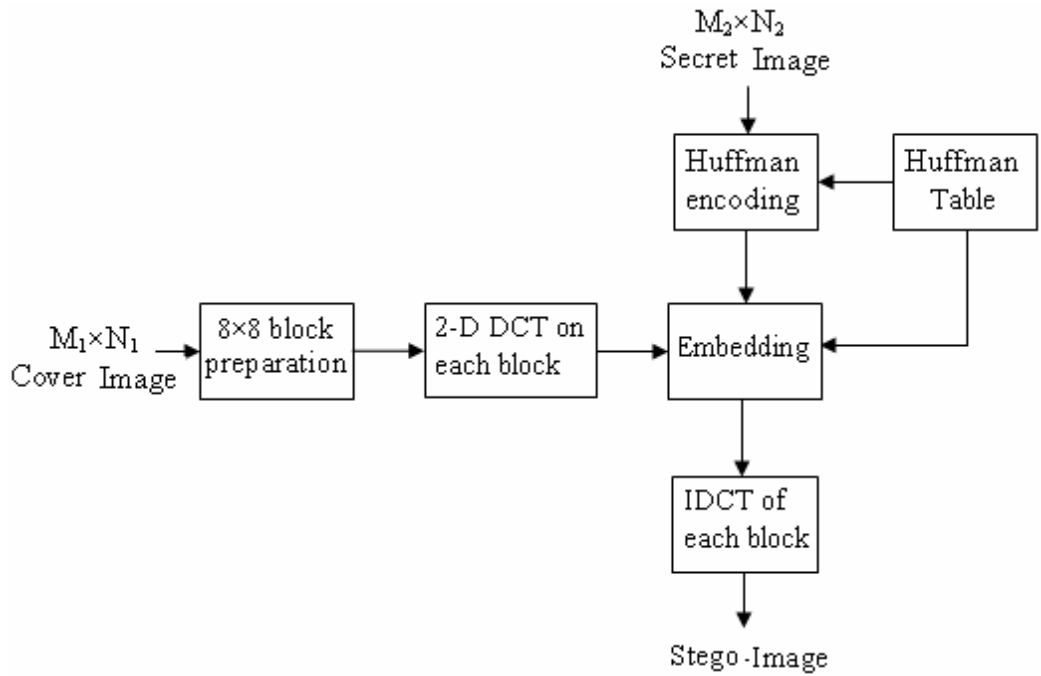

(a) Insertion of a Secret image (or message) into a Cover image

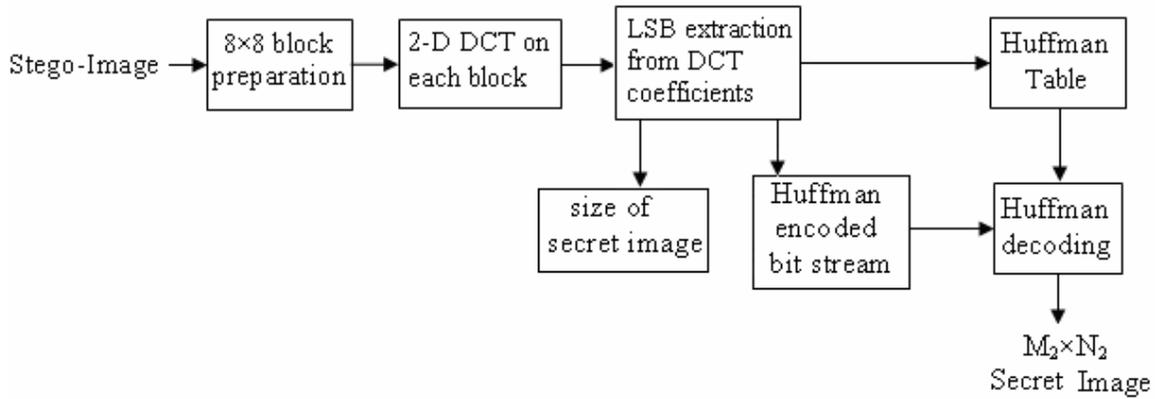

(b) Removal of Secret Image (or message)

**Figure 2:** Block diagram of the proposed steganography technique.

### 2.1 Discrete Cosine Transform

Let I(x,y) denote an 8-bit grayscale cover-image with x = 1,2,.......,$M_1$ and y = 1,2,.......,$N_1$. This $M_1 \times N_1$ cover-image is divided into 8 × 8 blocks and two-dimensional (2-D) DCT is performed on each of L = $M_1 \times N_1$ / 64 blocks. The mathematical definition of DCT is:

Forward DCT:

$$F(u,v) = \frac{1}{4} C(u)C(v) \sum_{x=0}^{7} \sum_{y=0}^{7} f(x,y) \cos\left[\frac{\pi(2x+1)u}{16}\right] \cos\left[\frac{\pi(2y+1)v}{16}\right]$$

for $u = 0,...,7$ and $v = 0,...,7$ (1)

$$\text{where } C(k) = \begin{cases} 1/\sqrt{2} & \text{for } k = 0 \\ 1 & \text{otherwise} \end{cases}$$

Inverse DCT :

$$f(x,y) = \frac{1}{4} \sum_{u=0}^{7} \sum_{v=0}^{7} C(u)C(v) F(u,v) \cos\left[\frac{\pi(2x+1)u}{16}\right] \cos\left[\frac{\pi(2y+1)v}{16}\right]$$

for $x = 0,...,7$ and $y = 0,...,7$ (2)

### 2.2 Huffman encoding and Huffman table ($H_T$)

Before embedding the secret image into cover image, it is first encoded using Huffman coding [24,25]. Huffman codes are optimal codes that map one symbol to one code word. For an image Huffman coding assigns a binary code to each intensity value of the image and a 2-D $M_2 \times N_2$ image is converted to a 1-D bits stream with length $L_H < M_2 \times N_2$. Huffman table ($H_T$) contains binary codes to each intensity value. Huffman table must be same in both the encoder and the decoder. Thus the Huffman table must be sent to the decoder along with the compressed image data.

### 2.3 8-bit block preparation

Huffman code H is decomposed into 8-bits blocks B. Let the length of Huffman encoded bits stream be $L_H$. Thus if $L_H$ is not divisible by 8, then last block contains r = $L_H$ % 8 number of bits (% is used as modulo operator).

### 2.4 Embedding of Secret Message / Image

We proposed the secret message/image embedding scheme comprises the following five steps:

**Step 1: DCT.**
Divide the carrier image into non overlapping blocks of size 8×8 and apply DCT on each of the blocks of the cover image f to obtain F using $eq^n$ (1).

**Step 2: Huffman encoding.**
Perform Huffman encoding on the 2-D secret image S of size $M_2 \times N_2$ to convert it into a 1-D bits stream H.

**Step 3:  8-bit block preparation.**
Huffman code H is decomposed into 8-bits blocks B.

**Step 4:  Bit replacement**
The least significant bit of all of the DCT coefficients inside 8×8 block is changed to a bit taken from each 8 bit block B from left to right. The method is as follows:

> For k=1 ; k≤1; k=k+1
>   LSB( ( F(u,v))$_2$) ← B(k) ;

Where B(k) is the $k^{th}$ bit from left to right of a block B and (F(u,v)$_2$ ) is the DCT coefficient in binary form.

**Step 5: IDCT.**
Perform the inverse block DCT on F using $eq^n$ (2)and obtain a new image $f_1$ which contains secret image.

**<u>Embedding Algorithm</u>**

**Input**: An $M_1$×$N_1$ carrier image and a secret message/image.
**Output**: A stego-image.

1. Obtain Huffman table of secret message/image.
2. Find the Huffman encoded binary bit stream of secret-image by applying Huffman encoding technique using Huffman table obtained in step 1.
3. Calculate size of encoded bit stream in bits.
4. Divide the carrier image into non overlapping blocks of size 8×8 and apply DCT on each of the blocks of the cover image.
5. Repeat for each bit obtained in step 3
   (a) Insert the bits into LSB position of each DCT coefficient of $1^{st}$ 8×8 block found in step 4.
6. Decompose the encoded bit stream of secret message/image obtained in step 2 into 1-D blocks of size 8 bits.
7. Repeat for each 8-bit blocks obtained in step 6
   (a) Change the LSB of each DCT coefficient of each 8×8 block(excluding the first) found in step 4 to a bit taken from left(LSB) to right(MSB) from each 8 bit block B.
8. Repeat for each bit of the Huffman table
   (a) Insert the bits into LSB position of each DCT coefficient
9. Apply inverse DCT using identical block size.
10. End.

**2.3 Extraction of the secret message / Image**

The stego-image is received in spatial domain. DCT is applied on the stego-image using the same block of size 8 × 8 to transform the stego-image from spatial domain to frequency domain. The size of the encoded bit stream and the encoded bit stream of secret message/image are extracted along with the Huffman table of the secret message/image. The block diagram of the extracting process is given in figure 1.1(b) and the extracting algorithm as follows:

**Extraction Algorithm**

**Input:** An $M_1 \times N_1$ Stego-image.
**Output:** Secret image.

1. Divide the stego-image into non overlapping blocks of size 8×8 and apply DCT on each of the blocks of the stego-image.
2. The size of the encoded bit stream is extracted from $1^{st}$ 8 × 8 DCT block by collecting the least significant bits of all of the DCT coefficients inside the $1^{st}$ 8×8 block.
3. The least significant bits of all of the DCT coefficients inside 8×8 block (excluding the first) are collected and added to a 1-D array.
4. Repeat step 3 until the size of the 1-D array becomes equal to the size extracted in step 2.
5. Construct the Huffman table by extracting the LSB of all of the DCT coefficients inside 8×8 blocks excluding first block and the block mentioned in step 3.
6. Decode the 1-D array obtained in step 3 using the Huffman table obtained in step 5.
7. End.

## 3. SIMULATION RESULTS

In this section, some experiments are carried out to prove the efficiency of the proposed scheme. The proposed method has been simulated using the MATLAB 7 program on Windows XP platform. A set of 8-bit grayscale images of size 512 × 512 are used as the cover-image to form the stego-image. The Figure 3 (a) – (d) shows the original cover (carrier) images and Figure 3 (e) shows the original secret message.

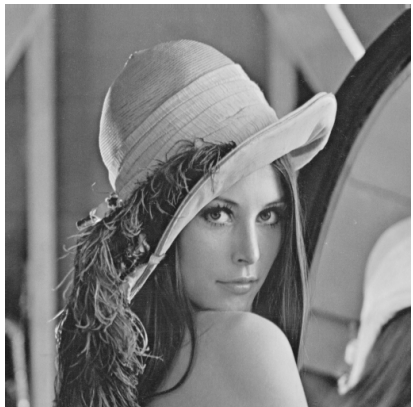
(a) Lenna

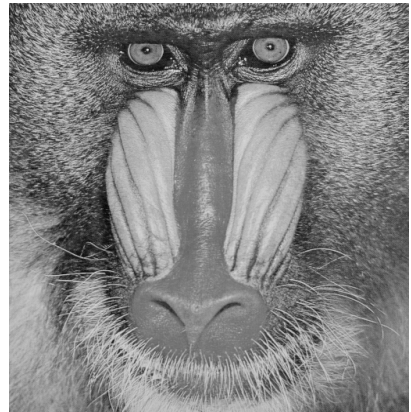
(b) Babban

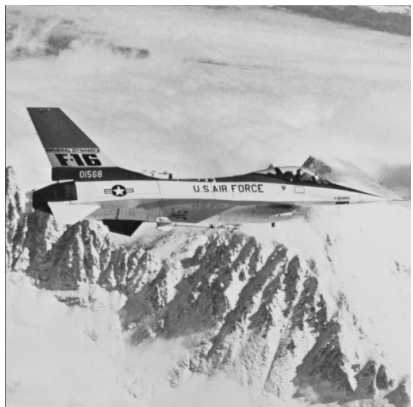
(c) Airplane

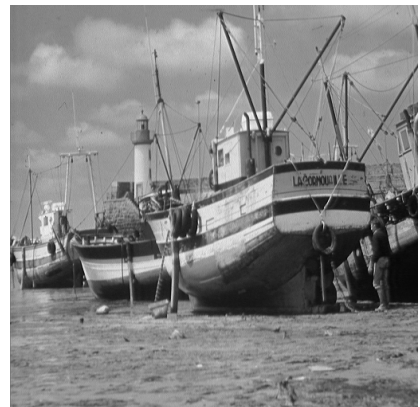
(d) Boat

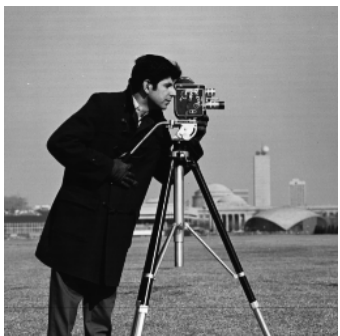

(e)

Figure 3: Four cover-images for simulations

The hiding capacity of secret messages/images is about $30 \times 10^4$ bits into a $512 \times 512$ cover (carrier) image. Here we are embedding a 8-bit grayscale image of size $192 \times 195$ into a 8-bit grayscale images of size $512 \times 512$ i.e. 294912 bits are embedded into a $512 \times 512$ carrier image. Table 1 exhibit the capacity and PSNR of four images.

**Table 1: Capacity and PSNR of four images**

| Images | Sise(in Kbyte) | Capacity(bits) | PSNR |
|--------|----------------|----------------|------|
| Lenna | 256 | 299520 | 50.48 |
| Baboon | 256 | 299520 | 50.28 |
| Airplane | 256 | 299520 | 50.91 |
| Boat | 256 | 299520 | 50.36 |

From table 1, it is observed that for all images, PSNR is greater than 50 and the hidden capacity is about 299520 bits. Table 2 shows that the hiding capacity and PSNR of our proposed algorithm is better than the one in reference [23], except only one case. Though the capacity of the image Baboon is better, but its PSNR is under 40dB.

**Table 2: PSNR comparison with the Modified Side Match Scheme (four-sided case) [10]**

| Cover Images | Modified Side Match | | | Proposed Method | | |
|--------------|------------|----------------|----------|------------|----------------|----------|
| | Size(Kbytes) | Capacity(bits) | PSNR(dB) | Size(Kbytes) | Capacity(bits) | PSNR(dB) |
| Lenna | 258 | 168,289 | 48.64 | 256 | 299520 | 50.48 |
| Baboon | 258 | 306,209 | 39.31 | 256 | 299520 | 50.28 |
| Airplane | 254 | 151,843 | 48.23 | 256 | 299520 | 50.91 |
| Boat | 256 | 207,497 | 44.33 | 256 | 299520 | 50.36 |

Figure 4((a),(b),(c) and (d)) shows the resulted stego-images of the proposed methods. Figure 4 (e) also shows the original secret message retrieved from the stego-images.

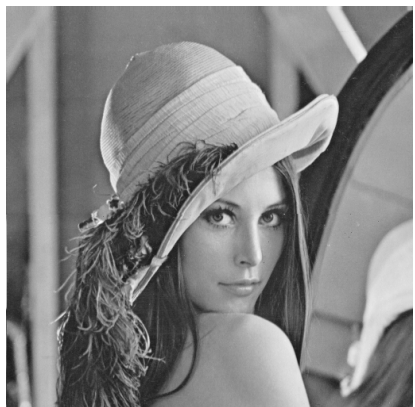

(a) Lenna

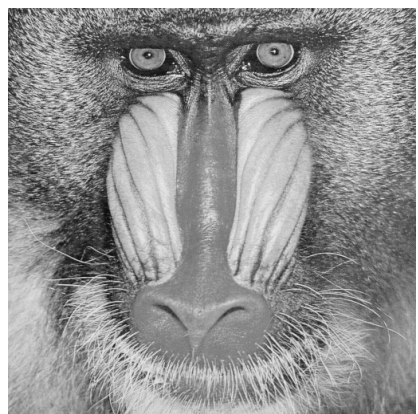

(b) Baboon

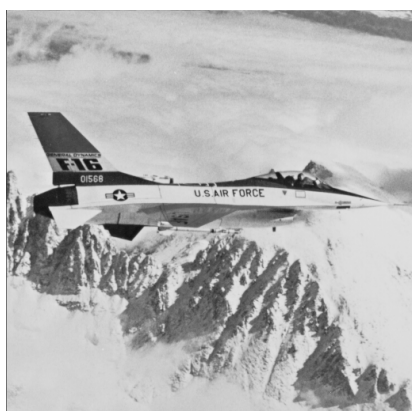

(c) Airplane

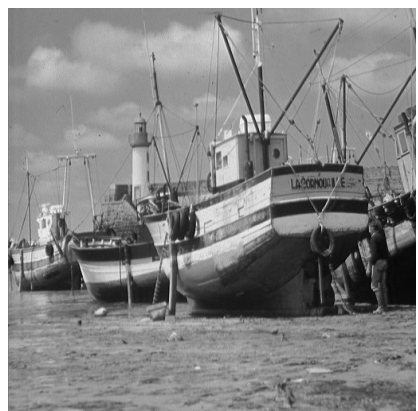

(d) Boat

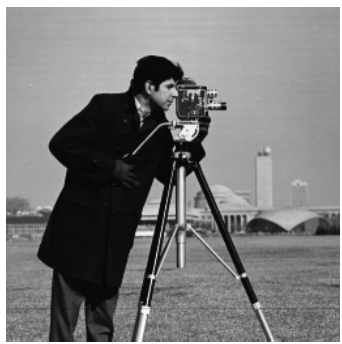

Figure 4: stego-images of the proposed methods

## 4. CONCLUSIONS

In this paper, we propose a steganography process in frequency domain to improve security and image quality compared to the existing algorithms which are normally done in spatial domain. According to the simulation results the stego-images of our method are almost identical to other methods' stego-images and it is difficult to differentiate between them and the original images. Our proposed algorithm also provides additional three layers of security by means of transformation (DCT and Inverse DCT) of cover image and Huffman encoding of secret image.

The demand of robustness in image steganography filed is not requested as strongly as it is in watermarking filed. As a result, image steganography method usually neglects the basic demand of robustness. In our proposed method the embedding process is hidden under the transformation i.e. DCT and inverse DCT. These operations and Huffman encoding of secret image keep the images away from stealing, destroying from unintended users and hence the proposed method may be more robust against brute force attack.